\documentclass{sig-alternate}
\newfont{\mycrnotice}{ptmr8t at 7pt}
\newfont{\myconfname}{ptmri8t at 7pt}
\let\crnotice\mycrnotice%
\let\confname\myconfname%

\usepackage[german,american]{babel}
\usepackage[T1]{fontenc}
\usepackage[utf8]{inputenc}
\usepackage{times}
\usepackage{textcomp}
\usepackage{graphicx}
\graphicspath{{}{./fig/}}

\usepackage{amssymb}
\usepackage{microtype} % for linebreaks in texttt

\usepackage{url}
\renewcommand{\tt}{\fontencoding{OT1}\fontfamily{cmtt}\selectfont}

\usepackage[usenames,dvipsnames]{color}
\usepackage{booktabs}
\usepackage[numbers,sectionbib]{natbib}%[round]
\makeatletter
\def\@copyrightspace{\relax}
\makeatother

\begin{document}

\title{Identifying Clickbait Posts on Social Media 
with an Ensemble of Linear Models}
\subtitle{The Carpetshark Clickbait Detector at the Clickbait Challenge 2017}

\numberofauthors{1}
\author{
  \alignauthor
  Alexey Grigorev\\
  \affaddr{OpenDataScience}\\
  \affaddr{contact@alexeygrigorev.com}\\
  \alignauthor
}

\maketitle

\begin{abstract}
The purpose of a clickbait is to make a link so appealing 
that people click on it. However, the content of such articles is 
often not related to the title, shows poor quality, and at the end
leaves the reader unsatisfied. 

To help the readers, the organizers of the clickbait 
challenge\footnote{\url{http://www.clickbait-challenge.org/}} asked the participants to build 
a machine learning model for scoring articles with respect to their
``clickbaitness''. 

In this paper we propose to solve the clickbait problem with an ensemble
of Linear SVM models, and our approach was tested successfully 
in the challenge: it showed great performance of 0.036 MSE and 
ranked 3rd among all the solutions to the contest. The code for the 
solution is available on 
GitHub\footnote{\url{https://github.com/alexeygrigorev/wsdmcup17-vandalism-detection}}.

\end{abstract}

%%%%%%%%%%%%%%%%%%%%%%%%%%%%%%%%%%%%%%%%%%%%%%%%%%%%%%%%%%%%%%%%%%%%%%%%
\section{Introduction}

Similar to \textit{baits} that fishermen use to catch fish, clickbaits
aim to catch the readers by getting their attention with an appealing title. 
These titles intent to spark the curiosity of the readers such that 
they want to follow the link. 

The following titles are good examples of clickbaits:

\begin{itemize}
\item Man tries to hug a lion. You won't believe what happens next!
\item 21 celebrities who ruined their faces with plastic surgeries
\item 15 hilarious tweets of stupid people
\end{itemize}

However, making the reader click the link and then displaying advertisements 
is often the sole purpose of the clickbaits~-- and the quality of the content 
is secondary for the publishers. The curiosity is usually not satisfied 
and the reader leaves the website. 

Social media websites such as Twitter or Facebook contain a large amount 
of clickbaits. It is possible to save the time of the readers by detecting 
if a title is clickbaiting and warning them, or even hiding the link 
altogether. This was the motivation of the clickbait challenge 
(\url{http://www.clickbait-challenge.org/}): the organizers
invited participants to develop a machine learning model for predicting 
if a post on social media is a clickbait. They prepared a large dataset 
of posts from Twitter with each post labeled as clickbaiting or not.

\subsection{Related Work}

It is not the first time the research community is concerned with the problem 
of detecting clickbaits. 

It is believed that Facebook downscores clickbaits in the user feeds, 
but there are no publications on their approach, only an announcement on their newsroom blog\footnote{\url{https://newsroom.fb.com/news/2014/08/news-feed-fyi-click-baiting/}}.

Use of Machine Learning to treat this problem was first reported by 
Potthast and others \cite{potthast:2016}, who collected a corpus of posts
from Twitter and used crowdsourcing to annotate them. 

To discover if their results can be improved, the same researches later 
organized the present challenge~-- the clickbait challenge \cite{potthast:2017a} on 
the TIRA platform \cite{potthast:2014}. In addition to the previous 
dataset, they gathered more data and released a significantly larger 
corpus of annotated posts from Twitter \cite{potthast:2017b}.

\section{Dataset and Challenge Description}

The provided dataset contains posts from a social media platform ``Twitter''.
This platform is often used by content providers to publish the links 
to their websites. Each post, ``tweet'', is a short message 
(up to 140 characters), which can be accompanied by a link and a picture. 

This is the information collected by the organizers: the content of 
each tweet, the link and the picture it contains. Each post is provided as a JSON object. For example:

\begin{verbatim}
{
  "id": "608999590243741697",
  "postTimestamp": 
      "Thu Jun 11 14:09:51 2015",
  "postText": 
      ["Some people are such food snobs"],
  "postMedia": 
      ["608999590243741697.png"],
  "targetTitle": 
      "Some people are such food snobs",
  "targetDescription": 
      "You'll never guess one...",
  "targetKeywords": 
      "food, foodfront, food waste...",
  "targetParagraphs": [
    "What a drag it is, eating kale ...",
    "A new study, published this ...", 
    "..."
  ],
  "targetCaptions": ["(Flikr/USDA)"]
} 
\end{verbatim}

The following information is available for each tweet:

\begin{itemize}
  \item \verb|postText|: the content of the tweet,
  \item \verb|postMedia|: the image that was posted alongside with the tweet (also made available by the organizers),
  \item \verb|targetTitle|: the title of the actual article,
  \item \verb|targetDescription| and \verb|targetKeywords|: the description and keywords from the meta tags of the article,
  \item \verb|targetParagraphs|: the actual content of the article,
  \item \verb|targetCaptions|: all captions in the article.
\end{itemize}

\begin{figure}
\centering
  \includegraphics[width=\columnwidth]{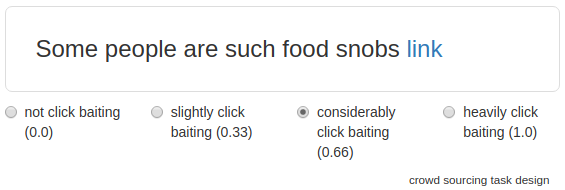}
  \caption{The design of the task evaluation. The screenshot is taken from 
\protect\url{http://clickbait-challenge.org/}.}
  \label{fig:eval-design}
\end{figure}

Each post is then assigned a ``clickbaitness'' score by human evaluators . 
It was possible to assign the post one of the following options (see fig.~\ref{fig:eval-design}):

\begin{itemize}
  \item not click baiting (0.0),
  \item slightly click baiting (0.33),
  \item considerably click baiting (0.66),
  \item heavily click baiting (1.0).
\end{itemize}

Multiple people evaluated each post, and all the responses were saved. 
For example, the post titled ``Apples' iOS 9 `App thinning' feature will
give your phone's storage a boost'' was evaluated by five people and the 
following scores were obtained:

\begin{itemize}
  \item 0.00 -- 3 times,
  \item 0.33 -- once,
  \item 0.66 -- once.
\end{itemize}

Then the mean score was calculated, which is 0.2 for this post. This score 
indicates that the post not clickbaiting. This is the target variable the 
contestants should predict: the mean clickbaitness score of each post. 
The solutions then were evaluated by Root Mean Squared metric~-- a common
metric for evaluating regression models.

\begin{table}
\centering
\begin{tabular}{|c|c|c|c|}
  \hline
  \textbf{Dataset} & \textbf{Posts} & \textbf{Clickbaits} & \textbf{Not Clickbaits} \\
  \hline
Training & 2495 & 762 & 1697 \\
Validation & 19538 & 4761 & 14777 \\
Unlabeled & 80012 & n/a & n/a \\
Testing & ? & ? & ? \\
  \hline
\end{tabular}
  \caption{The datasets provided by the organizers.}
  \label{tab:data}
\end{table}

Two labeled datasets were provided by the organizers: first, the 
dataset from the original
work by Potthast et al \cite{potthast:2016} with 2,495 posts: 762 clickbaits
and 1,697 not clickbaits. When the competition was already in progress,
the second dataset was released: it contained 19,538 posts with 
4,761 clickbaits and 14,777 not clickbaits \cite{potthast:2017b}.
The third labeled dataset was not provided: it was used for evaluating 
the models of the contestants.

In addition to the labeled dataset, another set of approximately 80,000 
tweets was available for use. The records in this dataset followed 
the same format, but no ground truth labels were collected for 
these posts (see table~\ref{tab:data}).

Finally, the contestants were free to use any external data source.

\section{Approach}

In this section we present our approach to the challenge in details.
First, we describe the hardware and software used for the solution,
then talk about external datasets we used for the challenge. After 
that we describe the features we extracted from the challenge dataset 
as well as the models built on these features. 

\subsection{Environment}

The experiments were performed on a Linux Ubuntu server with 32GB RAM and 8 Cores.

We used Python 3.4 and the PyData stack for our development:

\begin{itemize}
  \item \texttt{numpy} 1.12.1 for numerical operations \cite{van2011numpy};
  \item \texttt{scipy} 0.19.0 for storing sparse data matrices \cite{jones2015scipy};
  \item \texttt{pandas} 0.18.1 for tabular data manipulation \cite{mckinney2014pandas};
  \item \texttt{nltk} 3.2.1 for text manipulation \cite{bird2006nltk}, 
  \item \texttt{scikit-learn} 0.18.1 for data preprocessing and machine learning \cite{pedregosa2011scikit}.
\end{itemize}

We used Anaconda --~a distribution of Python with many
scientific libraries pre-installed\footnote{\url{https://www.continuum.io/downloads}},
including all the aforementioned libraries.

\subsection{External Data}

In the competition the participants were allowed to use 
any external data source. The provided labelled datasets are not very large,
which is why we decided to obtain additional data: 
we believed that it should help improve the quality of the resulting models. 

To obtain this additional data, we used the approach described in the 
post ``Identifying Clickbaits Using Machine Learning'' by 
A.~Thakur\footnote{\url{https://www.linkedin.com/pulse/identifying-clickbaits-using-machine-learning-abhishek-thakur/}}. 
Namely, we identified multiple Facebook groups that contained mostly clickbaiting
posts, and used the information there as positive labeled training instances.
Likewise, we picked up multiple groups with news which we deemed
not clickbaiting and assigned them the negative label. 
To retrieve the information from Facebook we used the Facebook 
scraper opensourced by 
M.~Wolf\footnote{\url{https://github.com/minimaxir/facebook-page-post-scraper}}.

This way we obtained the content of the following Facebook groups: 

\begin{itemize}
  \item clickbaiting (88.7k posts):
    \begin{itemize}
      \item buzzfeed (42.8k posts),
      \item clickhole (14k posts),
      \item upworthy (30k posts),
      \item StopClickBaitOfficial (1.9k posts).
    \end{itemize}
  \item not clickbaiting (154.9k posts):
    \begin{itemize}
      \item CNN (65.7k posts),
      \item Wikinews (2.8k posts),
      \item NYTimes (86.4k posts).
    \end{itemize}
\end{itemize}

In total we collected approximately 244.5k posts, and around 40\%
of them were clickbaiting. 

A similar approach was used by the author when developing a clickbait detection 
model at Searchmetrics, a company that is doing Search Engine Optimization.

\subsection{Modeling}

\textbf{Data Preparation and Features}

Each post in the provided dataset contains multiple text fields and a picture. 
Before using the text fields in the modeling, we applied to them the
following preprocessing procedure:

\begin{itemize}
\item remove repeating sentences,
\item remove HTML tags like \verb|<i>|, \verb|<b>| and so on,
\item remove English stop words,
\item extract the stem of each word with a Porter stemmer \cite{Manning2008IIR},
\item replace each number with a special token ``\verb|[n]|'',
\end{itemize}

The same preprocessing procedure was applied to the external dataset from Facebook. 

Then we used the standard Bag-of-Words approach to encode 
the textual information into a vector space \cite{Manning2008IIR}.
In additional to single tokens we also used 2- and 3-grams to 
capture the sequential nature of the textual data. 

We decided not to use image data as it seemed quite difficult and we 
believed it would not provide a significant improvement over the text-only 
solution.

\ \\

\textbf{Machine Learning models}

Our approach to the challenge was to train multiple linear models 
and then combine them with a tree-based model.

For model selection and validation we used the 5-fold cross-validation 
scheme, and the tables below report the mean score across all 5 folds.

\begin{table}
\centering
\begin{tabular}{|c|c|c|c|}
  \hline
  \textbf{Feature} & \textbf{MSE} & \textbf{Time, s} & \textbf{Best $C$} \\
  \hline
\verb|postText| & 0.039  & 1.7 & 0.1 \\
\verb|targetKeywords| & 0.060  & 1.3 & 0.5 \\
\verb|targetDescription| & 0.054  & 1.02 & 0.005 \\
\verb|targetCaption| & 0.053   & 1.7 & 0.001 \\
\verb|targetParagraphs| & 0.044  & 12.6 & 0.001 \\
\verb|targetTitle| & 0.047  & 1.7 & 0.01 \\
all concatenated & 0.042  & 16.3 & 0.001 \\
  \hline
\end{tabular}
  \caption{The performance of individual \texttt{LinearSVR} models for predicting the mean.}
  \label{tab:ind-mse}
\end{table}

\begin{table}
\centering
\begin{tabular}{|c|c|c|c|}
  \hline
  \textbf{Feature} & \textbf{MSE} & \textbf{Time, s} & \textbf{Best $C$} \\
  \hline
\verb|postText| & 0.014  & 0.7 & 0.01 \\
\verb|targetKeywords| & 0.014  & 0.3 & 0.01 \\
\verb|targetDescription| & 0.014  & 0.9 & 0.005 \\
\verb|targetCaption| & 0.014   & 2.0 & 0.001 \\
\verb|targetParagraphs| & 0.015  & 11.5 & 0.001 \\
\verb|targetTitle| & 0.014  & 1.0 & 0.01 \\
all concatenated & 0.015  & 14.6 & 0.001 \\
  \hline
\end{tabular}
  \caption{The performance of individual \texttt{LinearSVR} models for predicting the standard deviation.}
  \label{tab:ind-std}
\end{table}

For each text feature we trained an SVM regression model with a linear kernel
(\verb|LinearSVR| from scikit-learn, which is based on the LIBLINEAR 
library \cite{fan2008liblinear}). These sets of models were trained to predict 
the mean clickbaitness score of each post. 
The performance of each feature is reported in the table~\ref{tab:ind-mse}.
The best performing individual model is the model based on the post text~--
the actual message written on twitter. This model achieves 0.039 MSE 
on cross-validation.

Since our model is linear, it is possible to look at the coefficients to 
discover what are the strongest indicators of clickbaitness. For example, 
the following are the top 10 features with the largest positive weight
(note that the tokens are stemmed):

\begin{itemize}
\item \verb|[n] pictur|
\item \verb|[n] thing|
\item \verb|[n] artist|
\item \verb|[n] way|
\item \verb|[n] celebr|
\item \verb|here come|
\item \verb|shocker|
\item \verb|whoa|
\item \verb|[n] meme|
\item \verb|wat|
\end{itemize}

This suggests that the numbers (the ``\verb|[n]|'' token) in the post 
is a very good sign that the post is clickbaiting.

In addition to that
we trained another set of models which were used to predict the standard 
deviation of clickbaitness for each post (see table~\ref{tab:ind-std}).
For this target all features performed similarly achieving around 0.014 MSE.

Lastly, we trained a binary classification model
on our external data. For that we used SVM with a linear kernel,
\verb|LinearSVC| from scikit-learn, based on the LIBLINEAR 
library \cite{fan2008liblinear}.
To evaluate the performance there we used a hold-out dataset, and obtained 
AUC of 95\%. The trained model was then applied to the textual fields of the 
competition dataset, and the output was used an a set of additional features. 

\ \\

\textbf{Ensembling}

After training individual models we combined them using stacking \cite{wolpert1992stacked}:
the output of these models was used as input to another second-level model. 
The individual linear models were stacked using Extremely Randomized Trees \cite{geurts2006extremely}
(\verb|ExtraTreeRegressor| in scikit-learn)~-- a variation of the
Random Forest model that works especially good for stacking 
because it rarely overfits.

\begin{table}
\centering
\begin{tabular}{|c|c|c|c|}
  \hline
  \textbf{Features} & \textbf{MSE} \\
  \hline
Mean only & 0.0331 \\
Mean + Facebook & 0.0327 \\
Mean + std & 0.0326 \\
Mean + std + Facebook & 0.0326 \\
  \hline
\end{tabular}
  \caption{The performance of the ensemble models.}
  \label{tab:ensembles}
\end{table}

We trained the ensemble on different feature subsets, and established that 
the model that uses only mean and standard deviation features was the best one.
The model that also used the extra data from Facebook achieved 
the same performance, which is why we decided to use the simpler 
model without the Facebook features
(see table~\ref{tab:ensembles}). 

\begin{table}
\centering
\begin{tabular}{|c|c|c|c|}
  \hline
  \textbf{Feature} & \textbf{Importance} \\
  \hline
\verb|postText| mean & 0.32 \\
all concatenated mean  & 0.29 \\
\verb|targetTitle| mean   & 0.15 \\
\verb|postText| std  & 0.08 \\
\verb|targetKeywords| mean   & 0.07 \\
\verb|targetTitle| std    & 0.02 \\
all concatenated std   & 0.02 \\
\verb|targetKeywords| std    & 0.01 \\
\hline
\end{tabular}
  \caption{Feature importance: fraction of times the feature was used for a split in the tree.}
  \label{tab:importance}
\end{table}

According to the output of the ExtraTree model, the most important feature
is the \verb|postText| model for predicting the mean, 
which is not surprising, since it is best performing single 
model (see table~\ref{tab:importance}).

\section{Evaluation Results}

For the final evaluation we selected the best performing mean+std model and
executed it against the withheld testing data on the TIRA platform.

\begin{table}
\centering
\begin{tabular}{|c|c|c|}
  \hline
\textbf{Team name}  &   \textbf{MSE} &   \textbf{Running Time} \\
  \hline
zingel & 0.0332 & 00:03:27 \\
emperor & 0.0359 & 00:04:03 \\
\textbf{carpetshark} & 0.0362 & 00:08:05 \\
arowana & 0.0390 & 00:35:24 \\
pineapplefish & 0.0413 & 00:54:28 \\
  \hline
\end{tabular}
  \caption{Top 5 participants of the challenge.}
  \label{tab:final-standings}
\end{table}

Our approach took the 3rd position with 0.0362 MSE on the testing dataset
(see table~\ref{tab:final-standings}).

\section{Conclusion}

In this paper we approach the problem of identifying 
clickbaiting posts on social media. We show how to address the 
challenge with a small ensemble of linear models, and we conclude 
that the results are competitive: our model ranked high on the final 
standing.

%%%%%%%%%%%%%%%%%%%%%%%%%%%%%%%%%%%%%%%%%%%%%%%%%%%%%%%%%%%%%%%%%%%%%%%%
\begin{raggedright}
\bibliography{clickbait17-notebook-lit}
\end{raggedright}
\end{document}